\documentclass[]{article}
\usepackage{natbib}
\usepackage[applemac]{inputenc}
\usepackage[english]{babel}
\usepackage{epsfig}
\usepackage{graphicx}

\newcommand{\mean}[1]{\langle #1 \rangle}
\newcommand{\ud}{\mathrm{d}}
\newcommand{\der}[2]{\frac{\partial #1}{\partial #2}}
\newcommand{\Low}{\textbf{L}}
\newcommand{\Medium}{\textbf{M}}
\newcommand{\High}{\textbf{H}}

\begin{document}

\title{Very fine near-wall stuctures in scalar mixing}
\date{}
\author{Luca Galantucci \& Maurizio Quadrio \\ \\
Dip. Ingegneria Aerospaziale, Politecnico di Milano, Italy
}

\maketitle

\begin{abstract}
Passive scalar dynamics in wall-bounded turbulence is studied via Direct Numerical Simulations of plane channel flow, for a friction Reynolds number $Re_* = 160$ and a Schmidt number $Sc=1$. Peculiar to the present research is that the spatial resolution reaches far beyond what has been employed in similar past studies. Our aim is to examine the statistics of the most dissipative events across the various layers of the channel flow, and to compare them to the homogeneous isotropic case, where the recent studies by \cite{schumacher-sreenivasan-yeung-2005} and \cite{watanabe-gotoh-2007b} have described a range of scalar micro-scales that require extremely high spatial resolution to be properly resolved.

Resolution effects are observed on integral-scale quantities such as the mean profiles of the scalar dissipation and its variance. By examining probability distributions, it is found that the finest resolution is essential for correctly computing small-scale statistics in the near-wall region of the channel. As expected, this high-resolution requirement extends outwards to the channel centerline, where the behavior of isotropic turbulence is recovered. However, high-intensity scalar dissipation events are overemphasized by marginal resolution near the wall, while they are underemphasized in the central region.
\end{abstract}

\section{Introduction}
A passive scalar in a fluid flow is a diffusive physical quantity whose dynamical effects on the velocity field can be neglected. Examples of quantities that can often be conveniently treated as passive scalars are pollutants in liquids and gases, moisture in air, temperature in heated fluids (provided temperature fluctuations are sufficiently small), dye in liquids, chemical reactant species in industrial processes, etc. A comprehensive review of the literature on passive scalars and turbulence is given by \cite{warhaft-2000}.

The turbulent mixing of a passive scalar arises from two different yet concurring mechanisms: advective dispersion by large-scale motions, and small-scale diffusive molecular mixing. Experimental results \citep{warhaft-2000} as well as Direct Numerical Simulations  \citep{brethouwer-hunt-nieuwstadt-2003} show that these two physical processes are strongly coupled. As a consequence, a precise knowledge of the scalar's smallest scales is required for studying its integral-scale properties, such as the mean profile, the transport, the production and dissipation rates of scalar variance, and the scalar flux.

Understanding the smallest scales of scalar motion plays a key role in modeling the budget equation for the scalar variance, which is of interest in the LES and RANS approaches to turbulent flows. A satisfactory knowledge of the smallest scales might also improve our understanding of turbulence as a whole, as discussed by \cite{shraiman-siggia-2000}. Despite the linearity of the advection-diffusion equation with respect to the scalar, considerable internal intermittency and small-scale anisotropy characterize the passive scalar field \citep{sreenivasan-1991,  sreenivasan-antonia-1997}, and are larger than the ones of the corresponding turbulent velocity field \citep{chen-cao-1997, mydlarski-warhaft-1998, wang-chen-brasseur-1999, warhaft-2000, yeung-donzis-sreenivasan-2005}.

An effective tool for the numerical study of the smallest features of scalar motions is the Direct Numerical Simulation (DNS) of the Navier--Stokes equations. To obtain reliable and accurate results, however, an adequate spatial resolution must be chosen in order to resolve all the significant scales of motion.

The typical lengthscale of the smallest scalar motions is the mean Batchelor length scale $\mean{\eta_B}$, given by the mean Kolmogorov length scale $\mean{\eta}$ divided by the square root of the Schmidt number $Sc = \nu / \gamma$:
\begin{equation}
\mean{\eta_B} = \frac{\mean{\eta}}{\sqrt{Sc}} .
\label{eq:etab}
\end{equation}

In the expressions above, $\nu$ is the kinematic viscosity of the fluid and $\gamma$ is the scalar diffusivity; the operator $\mean{ \cdot }$ denotes an average taken over homogeneous directions and time. However, the fact that, on average, the smallest scales of the scalar motions are given by $\mean{\eta_B}$ does not imply that a spatial resolution set at $\mean{\eta_B}$ is sufficient. 

The proper spatial resolution for the DNS of passive scalars has been recently addressed by \cite{schumacher-sreenivasan-yeung-2005} and \cite{watanabe-gotoh-2007b} in the context of homogeneous and isotropic turbulence. The general conclusions of the two papers are similar: they point to the existence and dynamical relevance of scalar micro-scales that are misrepresented, should one set the spatial resolution of the numerical simulation based on $\mean{\eta_B}$. In fact, they demonstrate how the value of $\eta_B$ may become very small locally, and establish that in such cases scales larger than $\eta_B$ and smaller than $\mean{\eta_B}$ must be represented in the numerical simulation for the small-scale statistics to be correctly predicted.

In particular, \cite{schumacher-sreenivasan-yeung-2005} examined how these micro-scales in homogeneous turbulence behave as the Schmidt number is increased, whereas \cite{watanabe-gotoh-2007b} used the computing power of the Earth Simulator computer to carry out the same analysis at increasing Reynolds numbers with fixed $Sc=1$. The present paper deals with the same issue, and our original contribution will be to extend the analysis to wall-bounded turbulent flows. The plane channel flow is considered as the simplest prototypical wall flow that offers at the same time a simple geometry and the physical features of more complex wall-bounded flows. Aim of the paper is to show that the spatial resolution adopted in past DNS studies of wall turbulence with passive scalar does not resolve an important range of very small yet dynamically significant micro-scales, which are responsible for the extremely intermittent nature of the scalar field.

To demonstrate this statement, three DNS of the same channel flow at different spatial (and consequently temporal) resolutions are carried out. The lowest resolution corresponds roughly to the standard spatial resolution adopted in the past for most passive scalar studies: it is of the order of a few wall units, and comes from using for the passive scalar the spatial resolution typically employed for the velocity field. This spatial resolution is larger than $\mean{\eta_B}$. Our most resolved simulation, on the other hand, possesses a spatial resolution that goes well below the $\mean{\eta_B}$ level. The effects of varying resolution on the statistical features of the scalar field can thus be addressed, and their change with the wall distance described.

In particular, the focus of the present paper will be on the correct representation of the scalar dissipation. The dissipation $\epsilon_\theta$ of the scalar variance is defined as:
\begin{equation}
\epsilon_\theta = 2 \gamma \sum_{i=1}^3 \left(\der{\theta'}{x_i}\right)^2 ,
\label{eq:epstheta}
\end{equation}
where $\theta' \equiv \theta - \mean{\theta}$ is the fluctuation of the passive scalar about the local mean. Of course $\epsilon_\theta$ is particularly sensitive to the smallest scales of motion: it is known that $\epsilon_\theta$ possesses very fine spatial structure \citep{schumacher-sreenivasan-yeung-2005}, high internal intermittency, and significant small-scale anisotropy \citep{warhaft-2000}. In addition to being a suitable indicator of the smallest scalar motions, $\epsilon_\theta$ is significant in many physical processes which are central in both industrial and environmental fields. A correct modeling of both integral- and micro-scale scalar dissipation features can, for example, establish whether a chemical reaction occurs or not, and allows us to correctly predict quantitative features of environmental pollution aimed at establishing appropriate safety thresholds.

The outline of the paper is as follows. In \S\ref{sec:method} the characteristics of the numerical method will be briefly outlined, and its validation will be discussed in \S\ref{sec:validation} by reproducing the results of a channel flow DNS performed by \cite{johansson-wikstrom-1999}. In \S\ref{sec:parameters} the numerical parameters employed in the simulations will be introduced, with a view to comparing between the spatial resolutions used here and the typical resolution employed in the past. The main results will then be presented, illustrating first in \S\ref{sec:integral-quantities} the effects of the spatial resolution on integral-scale scalar quantities (mean profile and variance of scalar dissipation) and then in \S\ref{sec:scalar-dissipation} the effects of varying resolution on the micro-structural features of the scalar field. Finally, a brief summary and some conclusive remarks will be given in \S\ref{sec:conclusions}.

\section{The numerical method}
\label{sec:method}

The DNS code employed in this work has been developed from the pseudo-spectral, mixed-discretization, parallel algorithm introduced by \cite{luchini-quadrio-2006} for the DNS of the velocity field for wall-bounded turbulent flows. The extension of the original code to include the dynamics of a passive scalar has been implemented with the key requirement of keeping the same parallel strategy, and thus retaining the same computational efficiency. Since the wall-normal velocity -- wall-normal vorticity formulation is used for the momentum equations, the evolutive equation for the passive scalar:
\begin{equation}
\der{\theta}{t} + u_i\der{\theta}{x_i}=
\gamma \frac{\partial^2 \theta}{\partial x_i \partial x_i}
\label{eq:theta}
\end{equation}
is written to be formally identical to the wall-normal vorticity equation.

We indicate with $x$, $y$ and $z$ the streamwise, wall-normal and spanwise directions respectively. The corresponding velocity components are $u$, $v$ and $w$, and the passive scalar is $\theta$. The computational domain has extensions $L_x$, $L_y=2 \delta$ and $L_z$ in the corresponding directions. The friction Reynolds number $Re_*$ is based on the channel half-width $\delta$, on the fluid's kinematic viscosity $\nu$ and on the friction velocity $u_*$, and is defined as $Re_*= u_* \delta / \nu$. The scalar concentration is expressed through the Schmidt number $Sc = \nu / \gamma$.

A Fourier expansion is used in the homogeneous directions, whereas fourth-order accurate, explicit compact finite-difference schemes are used to compute derivatives in the $y$ direction. The number of discretization modes (points) is indicated with $N_x$, $N_y$ and $N_z$. The collocation points in the near-wall direction are smoothly stretched from the wall to the centerline. Full dealiasing is used in the homogeneous directions. The time-integration algorithm uses a third-order Runge-Kutta scheme for the computation of the convective non-linear terms and a second-order Crank-Nicholson scheme for the evaluation of the viscous-implicit terms.

The flow in the channel is made statistically stationary by a constant mean pressure gradient applied in the streamwise direction. Boundary conditions in the homogeneous directions are periodic whereas at the walls, no-slip and no-penetration conditions are imposed to the velocity field, and $\theta$ is set to a constant value. A mean scalar gradient is thus established between the two walls, ensuring a mean passive scalar profile in the wall-normal direction. This boundary condition for the passive scalar has been already employed for example by \cite{kim-moin-1989}, \cite{johansson-wikstrom-1999}, and \cite{kawamura-abe-shingai-2000}. It is worth mentioning here that an alternative boundary condition can be used, that consists in imposing a constant scalar flux at each wall. Such a boundary condition has been chosen, for example, by \cite{kasagi-tomita-kuroda-1992}, \cite{kasagi-ohtsubo-1993}, \cite{kawamura-ohsaka-abe-yamamoto-1998} and \cite{kawamura-abe-matsuo-1999}. The differences between the two boundary conditions have been addressed by \cite{kawamura-abe-shingai-2000}: they show that scalar statistics, such as the mean profile, the root-mean-square fluctuation and the turbulent flux, are basically unchanged in the near-wall region, while they differ in the core region due to the either zero or non-zero mean scalar gradient in the channel's mid plane.

\subsection{Validation}
\label{sec:validation}

The numerical code is validated by replicating the DNS of a turbulent channel flow with passive scalar carried out by \cite{johansson-wikstrom-1999}. This simulation is at $Re_*=265$ and $Sc=0.71$, with $L_x = 4 \pi \delta$ and $L_z = 5.5 \delta$, as in the original reference. The spatial resolution employed is $N_x=256$, $N_y=256$ and $N_z=192$. After reaching a statistically steady state, our numerical simulation is continued for an overall averaging time of 4000 viscous time units, and 60 statistically independent flow fields are written to disk for further analysis.

\begin{figure}
\centering
\include{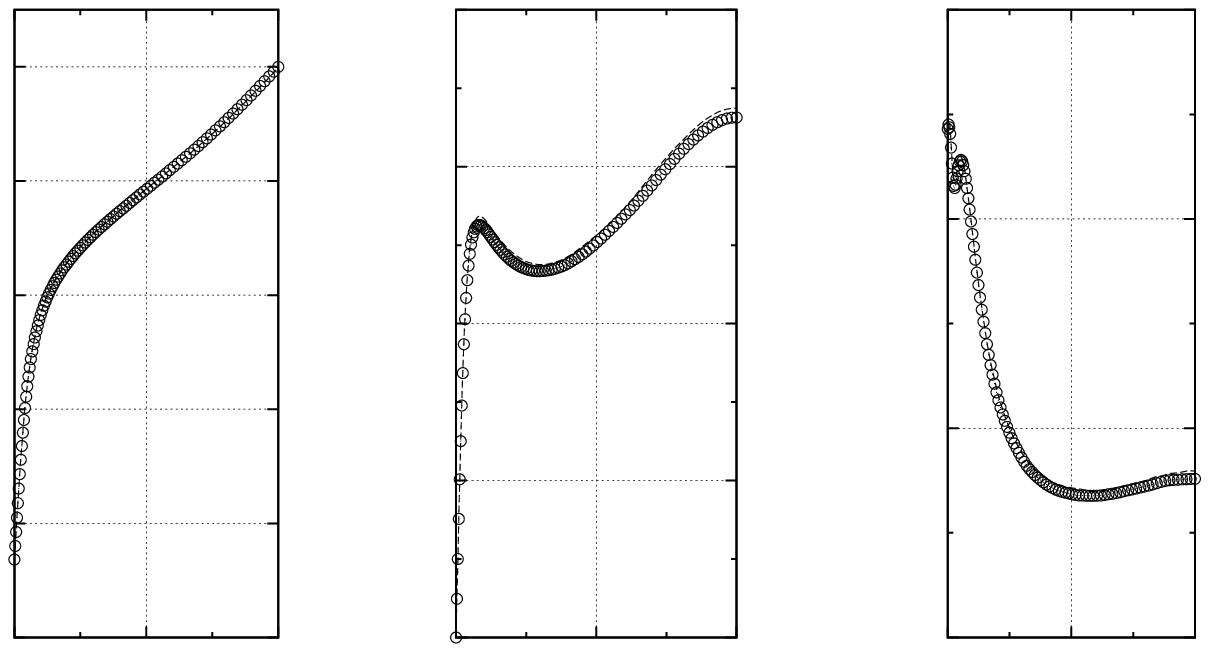}
\caption{Comparison between the present simulation (symbols) and results by \cite{johansson-wikstrom-1999} (dashed line) for $Re_*=265$ and $Sc=0.71$. All quantities in inner units. Left: wall-normal distribution of the mean passive scalar. Center: wall-normal profile of the r.m.s. value of the passive scalar fluctuations. Right: wall-normal mean profile of the dissipation of scalar variance. }
\label{fig:re265}
\end{figure}

Figure \ref{fig:re265} compares the results obtained by \cite{johansson-wikstrom-1999} to the present results. The two simulations are compared in terms of wall-normal profiles of the mean scalar $\mean{\theta}^+$, the root-mean-square value $\sigma_\theta^+$ of its fluctuations, and the mean scalar dissipation $\mean{\epsilon_\theta}^+$. The superscript $^+$ indicates a non-dimensional quantity after scaling with wall (inner) variables. The inner scaling quantities for the relevant physical variables involved in this work are as follows. The inner velocity scale is given by the friction velocity $u_*= (\tau_w/\rho)^{1/2}$, where $\tau_w$ and $\rho$ are respectively the wall-shear stress and the fluid density, and the inner time scale $t_*$ is defined by $t_* = \nu / u_*^2$. For the passive scalar field, the inner scale $\theta_*$ is given by:
\[
\theta_* = \frac{\gamma}{u_*} \frac{\ud \mean{\theta}}{\ud y}\bigg |_w ,
\]
where the subscript $w$ indicates a derivative evaluated at the wall.

Given the definition (\ref{eq:epstheta}) of the dissipation of scalar variance, its relevant scaling quantity $\epsilon_*$ is:
\[
\epsilon_* = \frac{\gamma}{Sc} \left( \frac{\ud \mean{\theta}}{\ud y}\bigg |_w \right)^2.
\]

A close inspection of figure \ref{fig:re265} establishes the agreement between the output of the present numerical tool and the results obtained by \cite{johansson-wikstrom-1999}. The collapse of the curves is such that only for the profile of $\sigma_\theta^+$ can the two simulations
be discerned. This little residual difference is most probably due to the different time span used for computing statistics.

\section{Computational parameters}
\label{sec:parameters}

The present work focuses on the smallest scales of motion, and thus the available computational resources are best spent on spatial resolution. As a consequence, our simulations carried out to study the passive scalar field in turbulent plane channel flow have neither particularly high values of $Re_*$ and $Sc$, nor particularly large sizes of the computational box. Its dimensions are set at $L_y = 2 \delta$, $L_x = 4.19 \delta$ and $L_z = 2.09 \delta$. The value of the friction Reynolds number is set at $Re_* = 160$. The value of the Schmidt number is set as $Sc = 1$, so that, according to Eq. (\ref{eq:etab}), the mean Kolmogorov and Batchelor lengthscales are identical.

\begin{table}
\begin{center}
\begin{tabular}{crrrcccccccc}
Simulation & $N_x$ & $N_z$ & $N_y$ & $\frac{\Delta x}{\mean{\eta_B}_w}$ &
$\frac{\Delta z}{\mean{\eta_B}_w}$ &
$\frac{\Delta y_{min}}{\mean{\eta_B}_w}$ &
$\frac{\Delta y_{max}}{\mean{\eta_B}_w}$ &
$\Delta x^+$ & $\Delta z^+$ & $\Delta y^+_{min}$ & $\Delta y^+_{max}$
\\ \hline \\
\Low ow &  64  & 64  &  128 & 6.54  &  3.28 &  0.54  & 2.47 & 10.46 & 5.25 & 0.86 & 3.95\\
\Medium edium & 340  & 170 &  128 & 1.23 &  1.23 & 0.54 &  2.47 & 1.97 & 1.97 & 0.86 & 3.95 \\
\High igh  & 680  & 340 &  256 & 0.62  &  0.62 &  0.27 & 1.24 & 1.00 & 1.00 & 0.43 & 1.98 \\
\end{tabular}
\end{center}
\caption{DNS parameters: grid spacings $\Delta x$, $\Delta y$, $\Delta z$ are expressed in terms of $\mean{\eta_B}_w$, the mean value of the Batchelor scale at the wall. In terms of viscous wall units, at the wall $\mean{\eta_B}_w^+=1.6$.}
\label{tab:parameters}
\end{table}

Three DNS at increasing spatial resolution are carried out. They are labelled with \Low, \Medium\ and \High\ throughout the paper, to indicate \Low ow, \Medium edium and \High igh resolution. The parameters defining the discretization of the three simulations are summarized in table \ref{tab:parameters}. The grid spacings are expressed, as usual in wall turbulence, in viscous wall units, but they are also reported in terms of $\mean{\eta_B}_w$, the mean Batchelor length scale evaluated at the wall.

The temporal discretization is adapted to the spatial discretization, so that the time step used in simulation \High\ is the finest. The overall averaging time (2400 viscous time units) and the number of statistically independent flow fields (60) stored on disk for further analysis for each of the three simulations are, however, left unchanged for all the simulations.

The spatial resolution of simulation \Low\ is comparable to the resolutions employed in most wall-turbulence DNS performed to date with passive scalars  \citep{kawamura-ohsaka-abe-yamamoto-1998, kawamura-abe-matsuo-1999, johansson-wikstrom-1999}. Such a spatial resolution is usually considered to be adequate as far as the velocity field is concerned \citep{moin-mahesh-1998}.

\begin{figure}
\include{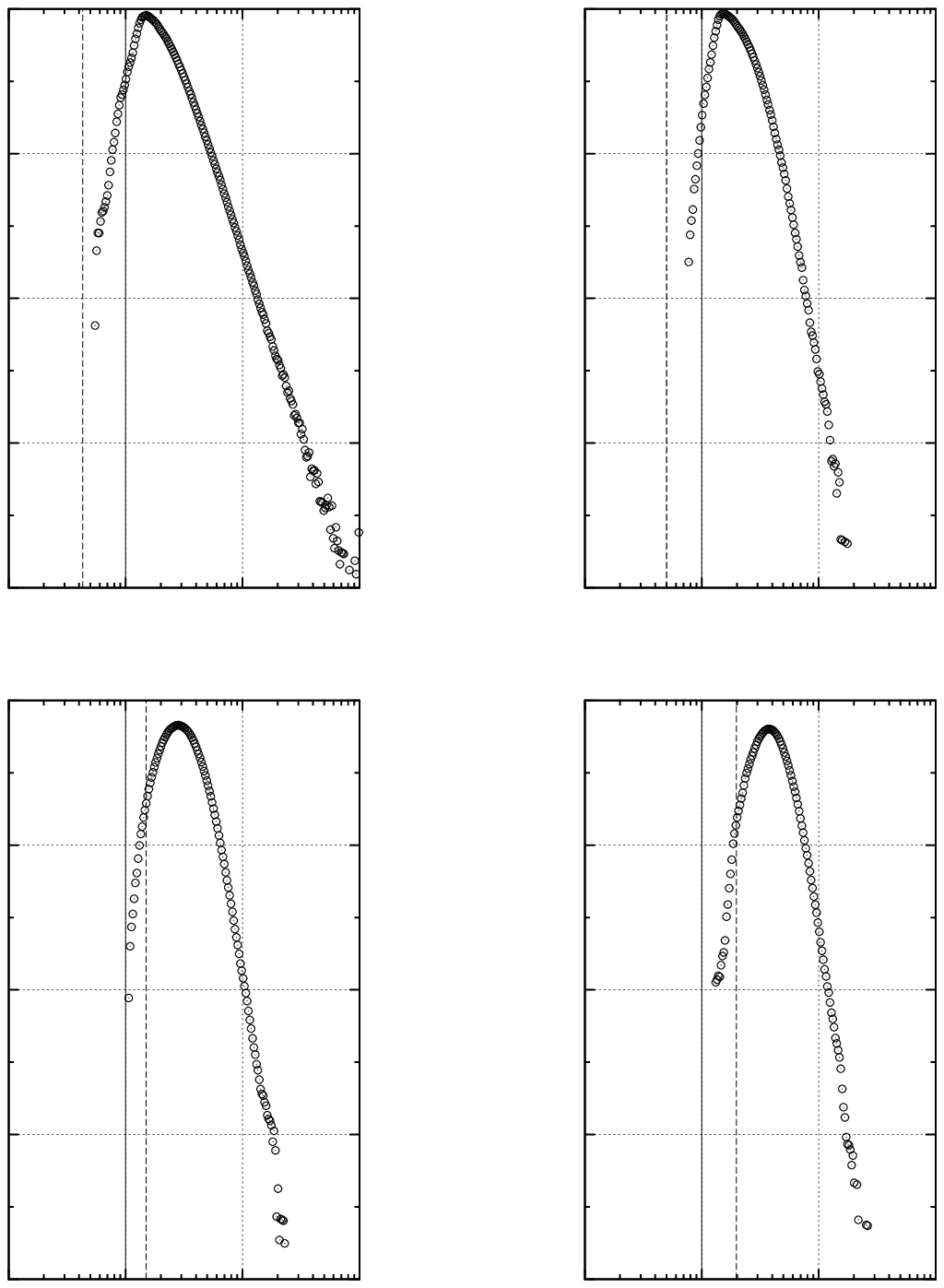}
\caption{Probability density functions of $\eta_B^+$ for: (a) $y^+=0$; (b) $y^+=5$; (c) $y^+=73$; (d) $y^+=160$. The vertical lines indicate the spatial resolution of simulation \High: the continuous vertical line is $\Delta x^+ = \Delta z^+ = 1$, whereas $\Delta y^+$  is shown as a dashed vertical line, and increases with $y$.}
\label{fig:pdf-etab}
\end{figure}

On the other hand, simulation \High\ has the highest resolution: owing to the increase with $y$ of the lengthscale $\mean{\eta}$, that equals $\mean{\eta_B}$ since $Sc=1$, the grid spacings $\Delta x$, $\Delta z$ and $\Delta y$ are consistently smaller than $\mean{\eta_B}$ in the whole wall-normal range. Still, motions at scales smaller than $\mean{\eta_B}$ do exist. Figure \ref{fig:pdf-etab} reports the statistical distribution of $\eta_B$ measured at a few wall-normal locations, and compares it to the spatial resolution of simulation \High\ at the same $y$ positions. It turns out that even the simulation employing the finest resolution could be only marginally resolved. At each wall-normal distance shown, in fact, either the streamwise (or spanwise) grid spacing, which is set constant to one wall unit throughout the channel, or the wall-normal spacing, that smoothly increases from the wall to the centerline, is larger than the smallest values locally assumed by $\eta_B$. 

The spatial resolution of simulation \Medium\ is midway between \Low\ and \High. Note that the streamwise and spanwise spatial resolution of case \Medium, when expressed in terms of  $\mean{\eta_B}_w$, is finer than the highest resolution employed in past DNS of passive scalar turbulent channel flows with $Sc\ge 1$ we are aware of. In absolute terms, this is the study recently carried out by \cite{schwertfirm-manhart-2007} whose finest grid spacings are $\Delta x^+=0.68$, $\Delta z^+=0.85$ and $0.18\le\Delta y^+\le0.75$. Their DNS, however, was aimed at investigating flows with high values of the Schmidt number, so that the resulting spatial resolution evaluated in terms of the Batchelor scale is rather coarse: $\Delta x / \mean{\eta_B}_w=3.09$, $\Delta z / \mean{\eta_B}_w = 3.86$ and $0.82 \le \Delta y / \mean{\eta_B}_w \le 3.41$.

\section{Statistical moments of $\epsilon_\theta$}
\label{sec:integral-quantities}

\begin{figure}
\centering
\include{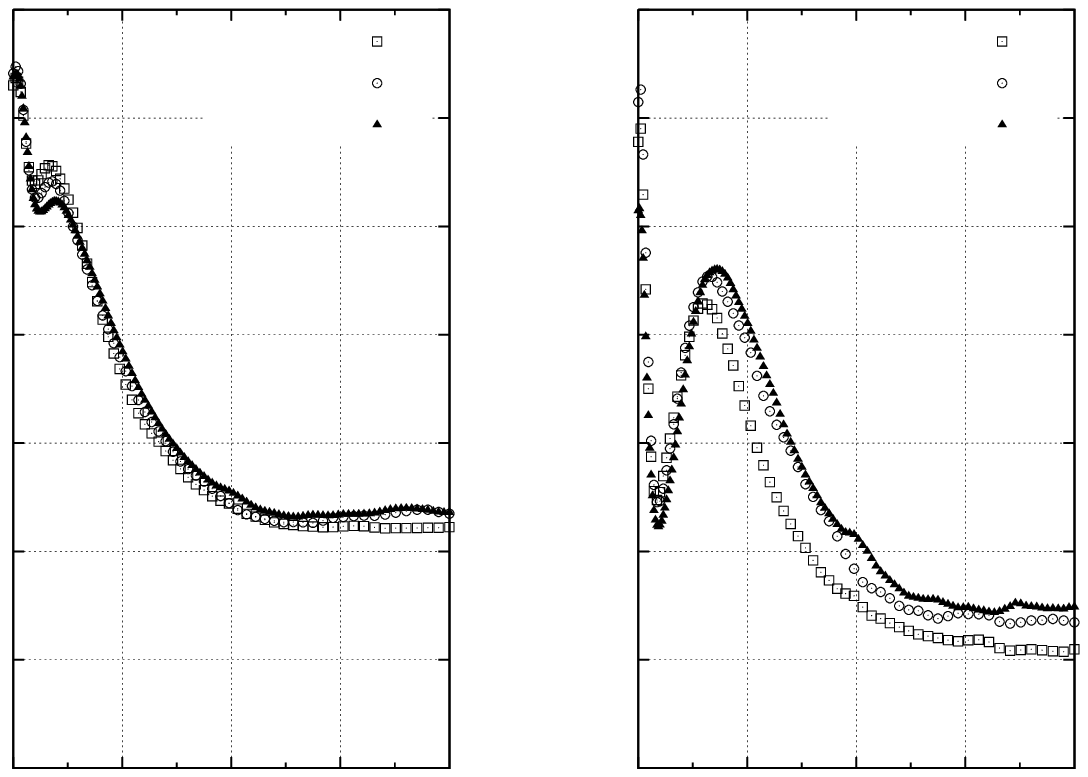}
\caption{Effects of the spatial resolution on the mean wall-normal distribution of the scalar dissipation rate $\epsilon_\theta$ (left) and on its variance $\sigma^2_{\epsilon_\theta}$ (right).}
\label{fig:eps-scalar}
\end{figure}

\begin{figure}
\centering
\include{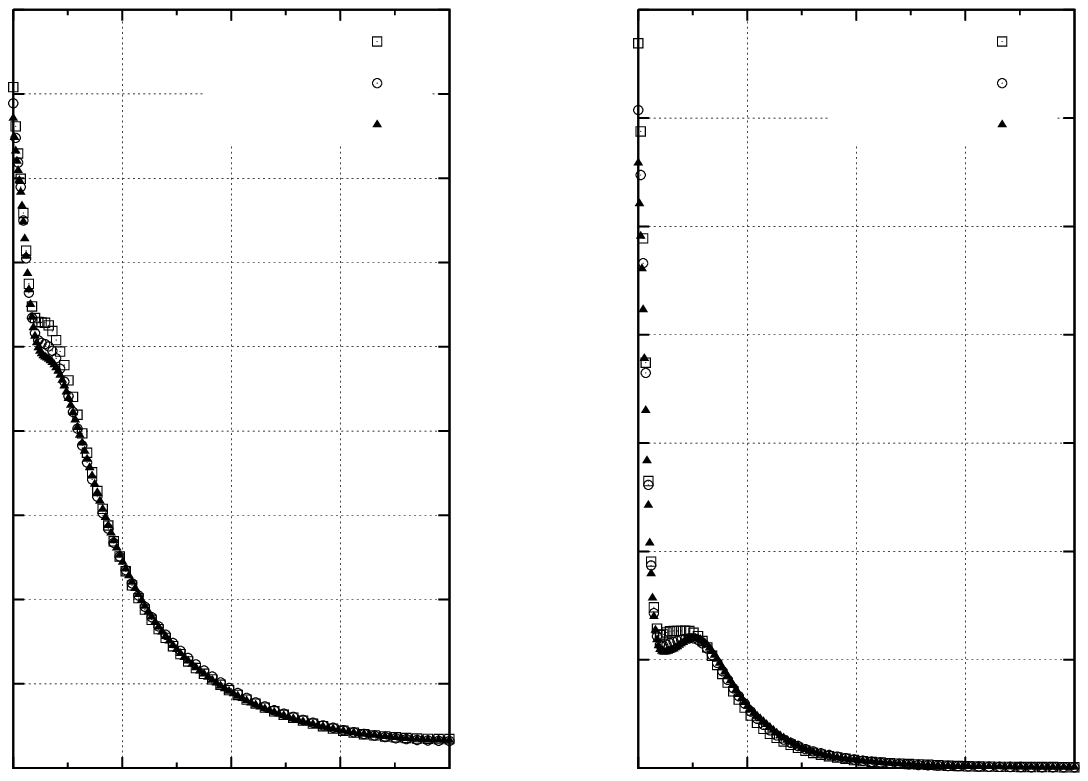}
\caption{Effects of the spatial resolution on the mean wall-normal distribution of the turbulent kinetic energy dissipation rate $\epsilon$ (left) and on its variance $\sigma^2_\epsilon$ (right).}
\label{fig:eps-velocity}
\end{figure}

The lowest-order statistical moments of the scalar dissipation $\epsilon_\theta$ are considered first. Figure \ref{fig:eps-scalar} shows the wall-normal distribution of statistical moments of $\epsilon_\theta$, respectively of order one (the mean profile) and order two (the variance), computed at different spatial resolutions. It can be easily appreciated how neither simulation \Low\ nor simulation \Medium\ have the spatial resolution required to compute these integral quantities in a mesh-independent way. Indeed, a complete overlap between curve \High\ and curve \Medium\ cannot be observed, and thus even simulation \High\ might still be slightly under resolved to correctly capture the full details of the fluctuations of $\epsilon_\theta$.

The three curves for $\mean{\epsilon_\theta}^+$ in figure \ref{fig:eps-scalar} (left) collapse in the central region of the channel and in the logarithmic layer, indicating that in these regions case \Low\ has enough resolution to represent $\mean{\epsilon_\theta}$ correctly. The three profiles are rather similar in the very-near-wall region too. Differences among the three curves become however evident in the buffer layer (say for $10 < y^+ < 30$), where the separation becomes significant and a change of about 10\% in the prediction of the mean value is observed. In this region, simulation \High\ predicts lower absolute values of $\mean{\epsilon_{\theta}}$, suggesting that on average the (absolute value of) dissipation of scalar variance associated to the finest scalar structures is lower.

Spatial resolution effects appear to be more relevant when the variance of $\epsilon_\theta$ is considered: in figure \ref{fig:eps-scalar} (right) the three curves remain separate for the whole $y^+$ range. The curves corresponding to cases \Low\ and \Medium\ present a similar behavior in the viscous sub-layer, suggesting that, at least in this flow region and at these resolution levels, the wall-normal spatial resolution, which is identical in simulations \Low\ and \Medium, could be the predominant factor.

It is important to note how the same analysis carried out for the dissipation rate $\epsilon$ of turbulent kinetic energy shows that these effects are peculiar to the $\epsilon_\theta$ field. Figure \ref{fig:eps-scalar} should be confronted with figure \ref{fig:eps-velocity}, that shows the wall-normal profiles of the first two statistical moments of $\epsilon$. Though slight resolution effects can still be observed, particularly between cases \Low\ and \Medium, it clearly emerges that the passive scalar field is characterized by a much finer and intermittent structure than the underlying velocity field \citep{warhaft-2000}. Hence turbulent flows with passive scalar, even at $Sc=1$ and $\mean{\eta_B} = \mean{\eta}$, present resolution requirements for correctly investigating integral-scale quantities of the scalar field which are more demanding than those required for the velocity field. When these requirements are not fully met, the insufficient resolution causes sizeable errors in the prediction of scalar-related mean quantities.

\section{PDF of $\epsilon_\theta$}
\label{sec:scalar-dissipation}

In this Section, the resolution effects already observed in terms of integral quantities are discussed in terms of their intensity distribution. We consider first in \S\ref{sec:scalardiss-high} particularly intense events, i.e. events characterized by extreme values of $\epsilon_\theta$, and examine how the spatial resolution affects the right tail of the probability density function of the quantity $(\epsilon_\theta - \mean{\epsilon_\theta}) / \sigma_{\epsilon_\theta}$. Then in \S\ref{sec:scalardiss-low} the same analysis is carried out for weak events, (i.e. events for which $\epsilon_\theta / \mean{\epsilon_\theta} \ll 1$) by looking at the left tail of the probability density function of $\epsilon_\theta / \mean{\epsilon_\theta}$.

\subsection{Strong dissipative events}
\label{sec:scalardiss-high}

\begin{figure}
\centering
\include{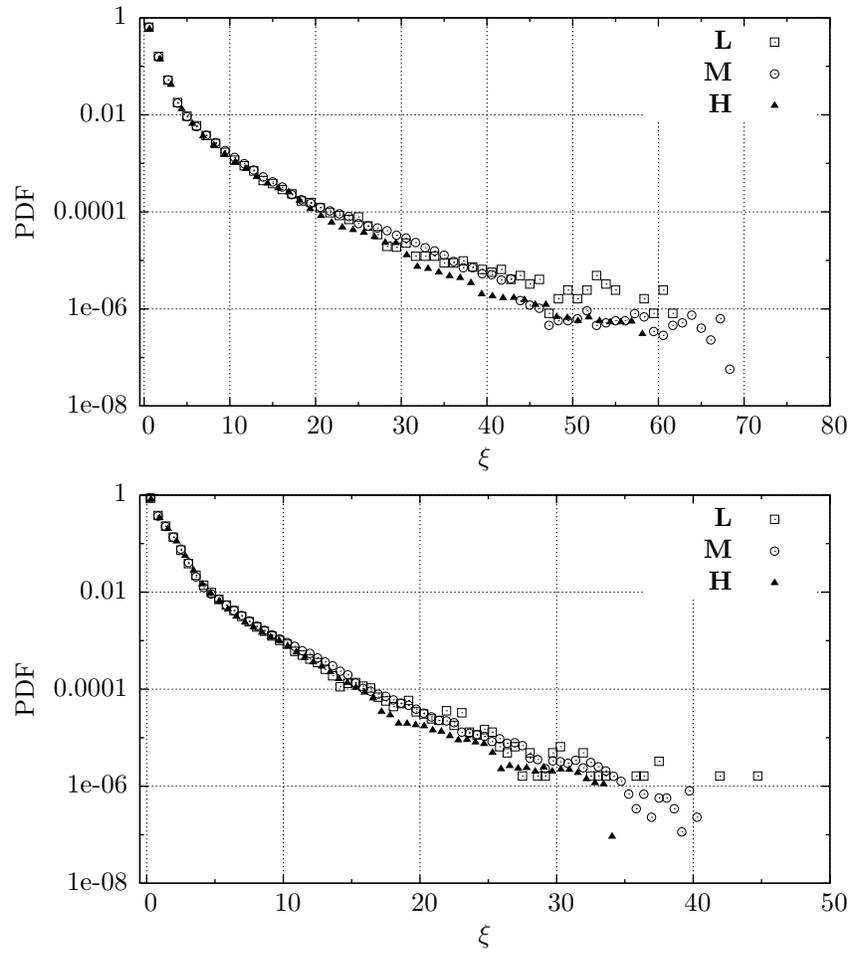}
\caption{PDF of $z = (\epsilon_\theta - \mean{\epsilon_\theta})/ \sigma_{\epsilon_\theta}$ for two wall-normal positions at $y^+=0$ (top) and $y^+=5$ (bottom).}
\label{fig:pdfstandard-nearwall}
\end{figure}

Figure \ref{fig:pdfstandard-nearwall} shows the probability density function of the quantity $z = (\epsilon_\theta - \mean{\epsilon_\theta})/ \sigma_{\epsilon_\theta}$ computed at $y^+=0$ (i.e. at the wall) and $y^+=5$ (in the viscous sublayer). Both the low-resolution simulations \Low\ and \Medium\ are characterized by PDF with wide tails, which appear to shrink when the spatial resolution is increased in case \High. This implies that, in the near-wall region, extremely intense events, that are present in cases \Low\ and \Medium\, disappear in simulation \High. This feature becomes more evident in the viscous sublayer, where the largest difference among the tails of the three curves is observed. The limited spatial resolution of case \Low\ also results in overestimating events of intermediate intensity. Overall, these dissimilarities account for the higher absolute value of $\mean{\epsilon_\theta}$ in the near-wall region previously observed in figure \ref{fig:eps-scalar} for cases \Low\ and \Medium.

\begin{figure}
\centering
\includegraphics[width=\textwidth]{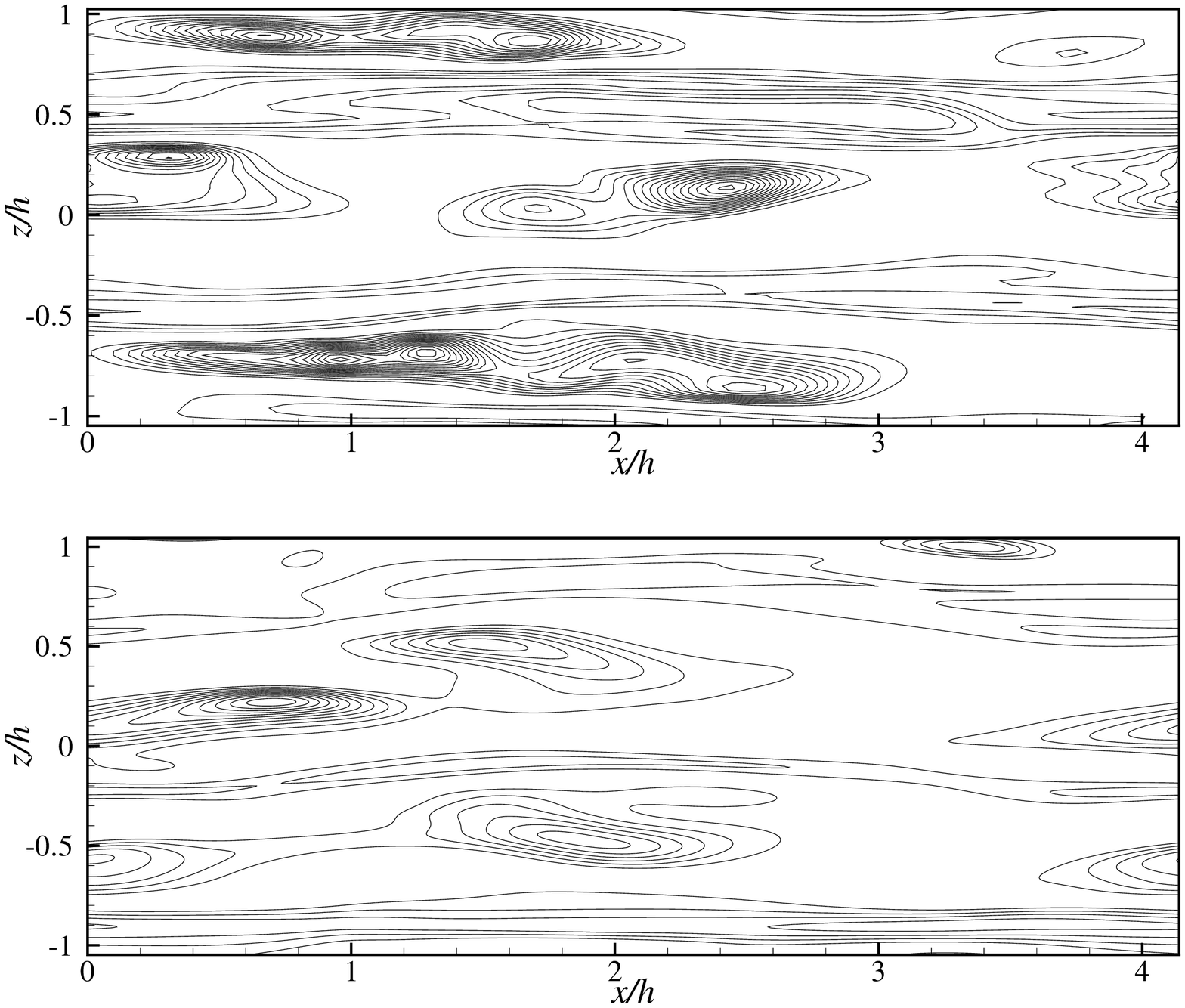}
\caption{Two-dimensional instantaneous wall-parallel cuts of the scalar dissipation field $\epsilon_\theta^+$ for $y^+=0$ for simulation \Low\ (top) and \High\ (bottom). Levels from 0 by 0.25 increments. Maxima are 4.15 (top) and 2.98 (bottom).}
\label{fig:cut-nearwall}
\end{figure}

These conclusions, based on statistical quantities, can be arrived at by also observing instantaneous snapshots of the $\epsilon_\theta$ field. In particular, a wall-parallel section of the computational domain taken at $y=0$ (i.e. at the wall) is shown in figure \ref{fig:cut-nearwall}, where the morphology of an $\epsilon_\theta$ field computed with the lowest (top) and the highest (bottom) resolutions are compared. The qualitative differences between the two slices help explaining the separation of the curves observed above in the context of figure \ref{fig:pdfstandard-nearwall}. Case \Low\ always shows a larger number of regions with intense $\epsilon_\theta$; their size is larger when compared to the analogous structures observed in case \High. Furthermore, the typical values of $\epsilon_\theta$ for events of extreme intensity appears to be larger in simulation \Low. In this pair of snapshots, for example, the instantaneous maximum of $\epsilon_\theta^+$ is 4.15 for case \Low\ and 2.98 for case \High. Analogous characteristics can be observed in wall-parallel cuts taken through the viscous sublayer. 

\begin{figure}
\centering
\include{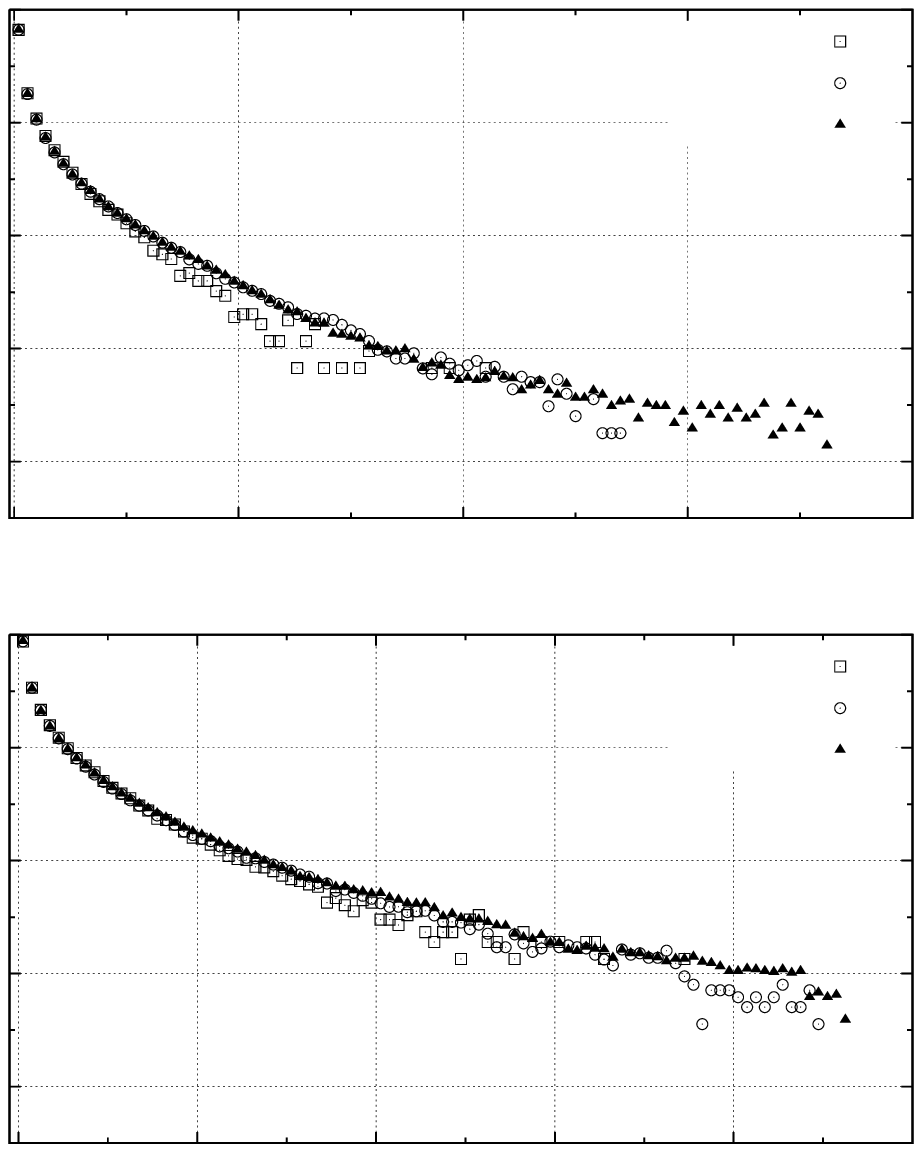}
\caption{PDF of $z = (\epsilon_\theta - \mean{\epsilon_\theta})/ \sigma_{\epsilon_\theta}$ for two wall-normal positions at $y^+=73$ (top) and $y^+=160$ (bottom).}
\label{fig:pdfstandard-loglayer}
\end{figure}

The pattern just described, i.e. one where marginal resolution tends to overemphasize extreme dissipations, reverses when the attention is shifted from the near-wall region to the logarithmic layer and the central region of the channel. Figure \ref{fig:pdfstandard-loglayer} reports again the PDF of $\epsilon_\theta$, this time computed at $y^+=73$ and $y^+=160$ (channel centerline). Here the most intense scalar dissipation events are reliably predicted by the highest spatial resolution only. This feature is more evident in the log-layer, where the tails of the probability density functions corresponding to cases \Medium\ and \High\ are definitely longer than the tail of the curve for case \Low. Moreover, in the log-layer, the intensity of the strongest events captured by the simulation increases with the adopted spatial resolution, as evident from the length of the PDF tails. This implies that resolution \Medium\ is certainly inadequate to capture the most intense scalar dissipation events in the log layer, and the same might be true for case \High\ too. At the channel centerline, the separation between the PDF tails becomes smaller, yet case \Low\ is evidently not capable of capturing the strongest events.

\begin{figure}
\centering
\includegraphics[width=\textwidth]{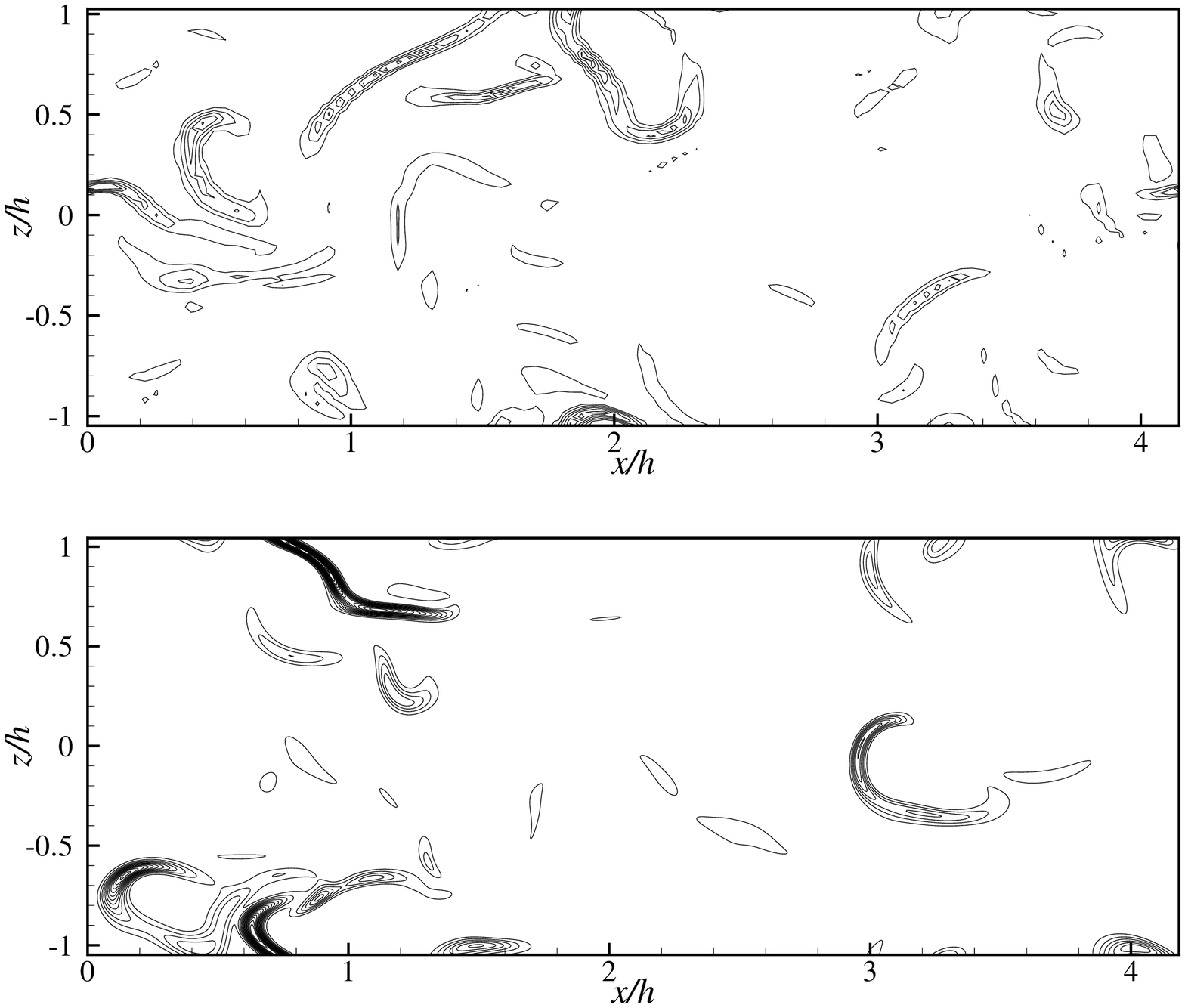}
\caption{Two-dimensional instantaneous wall-parallel cuts of the scalar dissipation field $\epsilon_\theta^+$ for $y^+=160$ for simulation \Low\ (top) and \High\ (bottom).
Levels from 0 by 0.25 increments. Maxima are 1.65 (top) and 4.35 (bottom).}
\label{fig:cut-centerline}
\end{figure}

Again, a look at instantaneous fields of $\epsilon_\theta$ in wall-parallel planes confirms these remarks. Figure \ref{fig:cut-centerline} shows sections at $y^+=160$ for simulations \Low\ and \High, and shows how the resolution of case \Low\ does not represent the $\epsilon_\theta$ field correctly, owing to its extremely intermittent spatial structure. The maxima of $\epsilon_\theta^+$ in these two particular slices are 1.65 (case \Low) and 4.35 (case \High).


\subsection{Weak dissipative events}
\label{sec:scalardiss-low}

\begin{figure}
\centering
\include{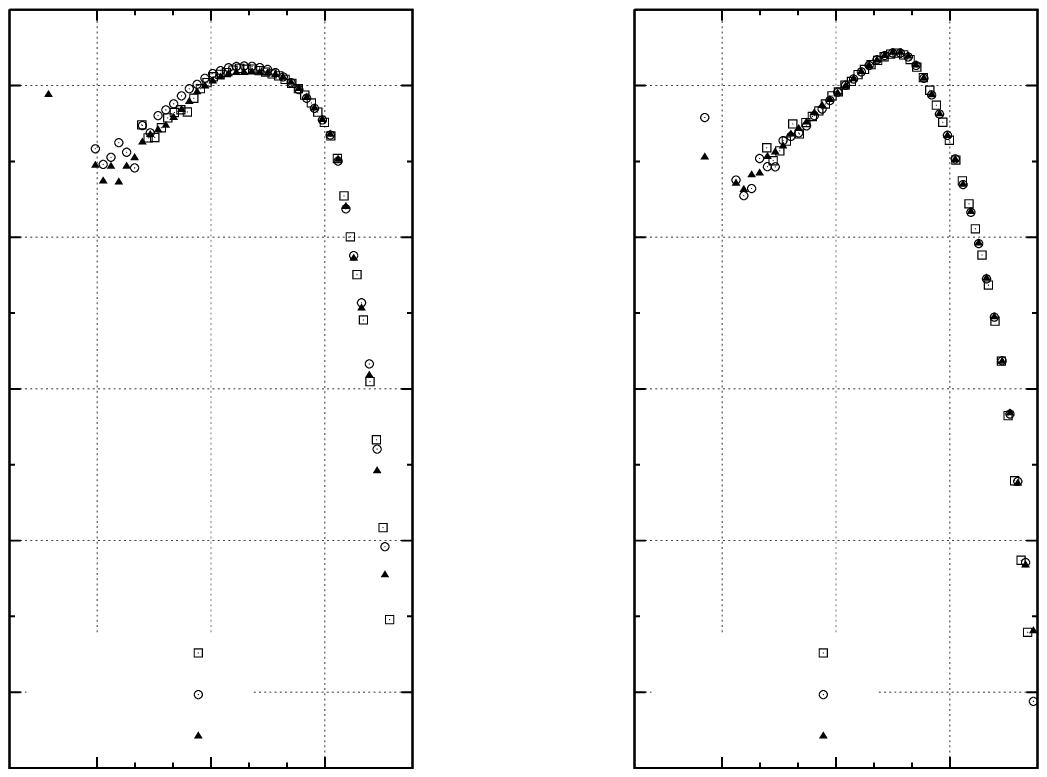}
\caption{PDF of $z = \epsilon_\theta / \mean{\epsilon_\theta}$ for 
$y^+=5$ (left), and $y^+=73$ (right).}
\label{fig:pdf-scalardiss}
\end{figure}

Figure \ref{fig:pdf-scalardiss} shows the PDFs of the quantity $z = \epsilon_\theta / \mean{\epsilon_\theta}$ for the three different simulations in the near-wall region and in the log-layer. The choice of this quantity, and the logarithmic scale used on the horizontal axis, emphasize the left side of the PDF, and highlight the effects of the spatial resolution on the weakest dissipation events.

The emerging pattern is that the left tail of the PDF for case \Low\ is consistently shorter than the tails for the other simulations. This means that a marginal spatial resolution is not capable to identify extremely weak events of scalar dissipation. As already observed by \cite{schumacher-sreenivasan-yeung-2005}, this can be explained by the fact that numerical simulations with poor resolution are characterized by relatively high noise (when compared to more resolved simulations) and this implies a lower signal-to-noise ratio. The noise becomes particularly significant in the regions of low-magnitude $\epsilon_\theta$ events, that end up being covered by the noise floor.

Simulations \Medium\ and \High, on the other hand, show no dissimilarities as far as the weakest $\epsilon_\theta$ events are considered, if exception is made for the viscous sub-layer (see again figure \ref{fig:pdf-scalardiss}), where the highest resolution is needed to observe the smallest events.

Looking at the same quantities in the central region of the channel (not shown) confirms these observations, whereas at the wall, on the opposite, the three curves essentially coincide at small $z$. This last observation is explained both by the isoscalar boundary condition employed in the present work (which implies a tendential uniformity of the scalar in wall-parallel planes for very small wall distances) and by the larger characteristic length scale of the structures in the near-wall region, which can therefore be represented with lower spatial resolution.

\begin{figure}
\centering
\includegraphics[width=\textwidth]{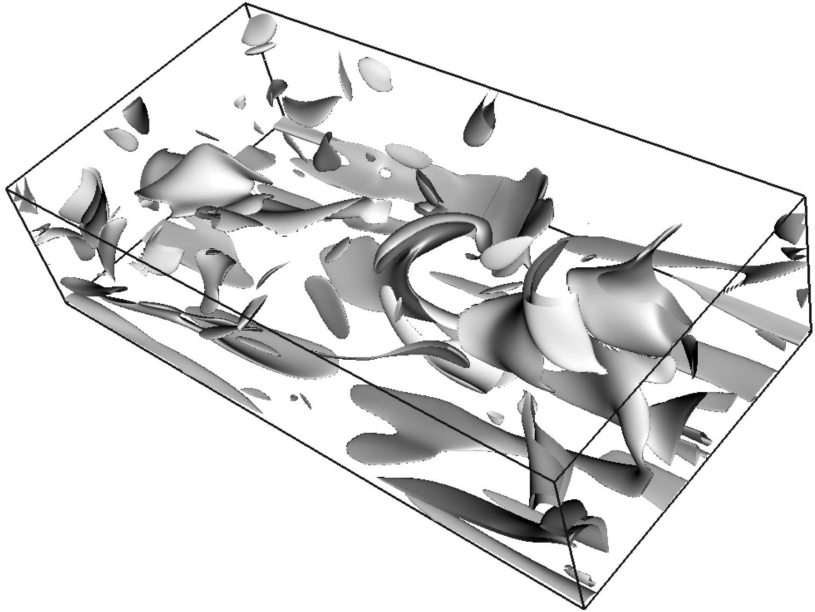}
\caption{Three dimensional visualization of iso-scalar dissipation surfaces corresponding to $\epsilon_{\theta}/\epsilon^*=0.73$ in the bottom half of the channel. }
\label{fig:3D}
\end{figure}

\section{Discussion and conclusions}
\label{sec:conclusions}

In the present work three Direct Numerical Simulations of a passive scalar in turbulent channel flow have been carried out at different spatial resolutions. The aim was to investigate the effects of the spatial resolution on the statistical features of the flow, and to describe their change with the distance from the wall. We have centered our analysis on the statistical description of the scalar dissipation $\epsilon_\theta$. Grid-dependent results have been observed not only in the statistical description of the passive scalar's micro-structure, but -- more surprisingly -- also in its integral-scale characteristics, unless the grid spacing in all the three directions is smaller than the local value of the Batchelor length scale.

The wall-normal profiles of $\mean{\epsilon_\theta}$ and its variance show that, due to the strong coupling between large-scale advective motions and small-scale molecular diffusive mixing, the most resolved DNS is needed even for the correct description of such integral-scale quantities. The required resolution goes far beyond  what has been routinely adopted in the past for similar simulations. The resolution requirements become of course more stringent as soon as the order of the considered statistical moments increases. Even though such a specific study has probably never been carried out for the velocity field, a comparison between the wall-normal mean profiles of $\epsilon_\theta$ and its variance with the corresponding profiles for the dissipation rate of turbulent kinetic energy strongly suggests that such strict requirements concern primarily the passive scalar, owing to its distinctive intermittent character.

Extra-fine spatial resolution becomes essential when the smallest scales of the passive scalar are the main feature one is interested in. The PDF of $\epsilon_\theta$ fluctuations, measured in the channel's log-layer and the mid plane and reported in \S\ref{sec:scalardiss-high} and \S\ref{sec:scalardiss-low}, do in fact show that the range of scales which are not resolved by standard spatial resolutions is responsible for the extremely intermittent nature of the passive scalar field, that is observed for both very weak and very intense scalar dissipation events. This is coherent with the pattern described by \cite{schumacher-sreenivasan-yeung-2005} and \cite{watanabe-gotoh-2007b} for homogeneous isotropic turbulence. When the wall is approached, on the other hand, the influence of the solid boundary reverses the picture. In this region of the flow we have shown how the incorrect representation of the smallest scales leads to a significant overestimate of the strongest $\epsilon_\theta$ events.

The varying effect of insufficient spatial resolution in the different layers of the channel flow is related to the changes of the $\epsilon_\theta$ field throughout the channel. The relatively large, high-dissipation structures which can be observed in near-wall horizontal sections of the flow field, visualized in figure \ref{fig:cut-nearwall}, do disappear, in fact, as the distance from the wall increases, and are replaced by thin and elongated structures, shown in figure \ref{fig:cut-centerline}, which recall, as it could have been expected, the structures observed by \cite{brethouwer-hunt-nieuwstadt-2003} and \cite{schumacher-sreenivasan-yeung-2005} in homogeneous and isotropic turbulence. According to these papers, the thin and elongated structures are twisted, folded, very close and almost parallel to each other, and are most likely cross-sections of the sheet-like structures of $\epsilon_\theta$ which can be observed in the three-dimensional visualization of the scalar dissipation field plotted in figure \ref{fig:3D}. The sheet-like structures are predominantly parallel to the wall in the near-wall region, and assume more of an isotropic character in the central region of the channel. The decreasing characteristic dimensions of high-$\epsilon_\theta$ structures for increasing distances from the wall explains why intense $\epsilon_\theta$ events are underestimated by simulation \Low\ in the channel's log-layer and near the mid plane of the channel.

The (at least partial) grid-dependency presented by several results discussed throughout this paper implies that even the highest spatial resolution employed (simulation \High) could be only marginally adequate. This is also suggested by the comparison, carried out in figure \ref{fig:pdf-etab}, between the statistical distribution of the Batchelor's scale $\eta_B$ and the spatial resolution of case \High\ in the different layers of the channel flow. Grid resolutions finer than the smallest local value of $\eta_B$ will therefore be required to investigate small-scale features of passive scalar mixing in wall-bounded turbulent flows.

\section*{Acknowledgements}
A. Johansson is gratefully acknowledged for having shared with us his data. LG has been partially supported by a PRIN 2005 grant on {\em Large-scale structures and wall turbulence}. The use of computer time on the system run by prof. P.Luchini at Universit\`a di Salerno is gratefully acknowledged.

\bibliographystyle{jfm}
\bibliography{../../mq}

\begin{thebibliography}{21}
\expandafter\ifx\csname natexlab\endcsname\relax\def\natexlab#1{#1}\fi

\bibitem[Brethouwer {\em et~al.\/}(2003)Brethouwer, Hunt \&
  Nieuwstadt]{brethouwer-hunt-nieuwstadt-2003}
{\sc Brethouwer, G., Hunt, J.C.R. \& Nieuwstadt, F.T.M} 2003 Micro-structure
  and {L}agrangian statistics of the scalar field with a mean gradient in
  isotropic turbulence. {\em J. Fluid Mech.\/} {\bf 474}, 193--225.

\bibitem[Chen \& Cao(1997)]{chen-cao-1997}
{\sc Chen, S. \& Cao, N.} 1997 Anomalous scaling and structure instability in
  three-dimensional passive scalar turbulence. {\em Phys. Rev. Lett.\/} {\bf
  78}~(18), 3459.

\bibitem[Johansson \& Wikstr\"om(1999)]{johansson-wikstrom-1999}
{\sc Johansson, A.~V. \& Wikstr\"om, P.M.} 1999 D{NS} and modelling of passive
  scalar transport in turbulent channel flow with a focus on scalar dissipation
  modelling. {\em Flow Turbulence and Combustion\/} {\bf 63}, 223--245.

\bibitem[Kasagi \& Ohtsubo(1993)]{kasagi-ohtsubo-1993}
{\sc Kasagi, N. \& Ohtsubo, Y.} 1993 {\em Direct numerical simulation of
  passive scalar field in a turbulent channel flow\/}, {\em Turbulent Shear
  Flow\/}, vol. VIII, pp. 97--119. Springer.

\bibitem[Kasagi {\em et~al.\/}(1992)Kasagi, Tomita \&
  Kuroda]{kasagi-tomita-kuroda-1992}
{\sc Kasagi, N., Tomita, Y. \& Kuroda, A.} 1992 Direct numerical simulation of
  passive scalar field in a turbulent channel flow. {\em Trans. ASME\/} {\bf
  114}, 598--606.

\bibitem[Kawamura {\em et~al.\/}(1999)Kawamura, Abe \&
  Matsuo]{kawamura-abe-matsuo-1999}
{\sc Kawamura, H., Abe, H. \& Matsuo, Y.} 1999 D{NS} of turbulent heat transfer
  in channel flow with respect to {R}eynolds and {P}randtl number effects. {\em
  Int. J. Heat Fluid Flow\/} {\bf 20}, 196--207.

\bibitem[Kawamura {\em et~al.\/}(2000)Kawamura, Abe \&
  Shingai]{kawamura-abe-shingai-2000}
{\sc Kawamura, H., Abe, H. \& Shingai, K.} 2000 D{NS} of turbulence and heat
  transport in a channel flow with different {R}eynolds and {P}randtl numbers
  and boundary conditions. 3rd Int. Symp. on Turbulence, Heat and Mass
  Transfer.

\bibitem[Kawamura {\em et~al.\/}(1998)Kawamura, Ohsaka, Abe \&
  Yamamoto]{kawamura-ohsaka-abe-yamamoto-1998}
{\sc Kawamura, H., Ohsaka, K., Abe, H. \& Yamamoto, K.} 1998 D{NS} of turbulent
  heat transfer in channel flow with low to medium-high {P}randtl number fluid.
  {\em Int. J. Heat Fluid Flow\/} {\bf 19}, 482--491.

\bibitem[Kim \& Moin(1989)]{kim-moin-1989}
{\sc Kim, J. \& Moin, P.} 1989 {\em Transport of passive scalars in a turbulent
  channel flow\/}, pp. 85--96. {\em Turbulent Shear Flows\/} VI. Springer.

\bibitem[Luchini \& Quadrio(2006)]{luchini-quadrio-2006}
{\sc Luchini, P. \& Quadrio, M.} 2006 A low-cost parallel implementation of
  direct numerical simulation of wall turbulence. {\em J. Comp. Phys.\/} {\bf
  211}~(2), 551--571.

\bibitem[Moin \& Mahesh(1998)]{moin-mahesh-1998}
{\sc Moin, P. \& Mahesh, K.} 1998 Direct numerical simulation: A tool in
  turbulence research. {\em Ann. Rev. Fluid Mech.\/} {\bf 30}, 539--578.

\bibitem[Mydlarski \& Warhaft(1998)]{mydlarski-warhaft-1998}
{\sc Mydlarski, L. \& Warhaft, Z.} 1998 Passive scalar statistics in
  high-{P}eclet-number grid turbulence. {\em J. Fluid Mech.\/} {\bf 358},
  135--175.

\bibitem[Schumacher {\em et~al.\/}(2005)Schumacher, Sreenivasan \&
  Yeung]{schumacher-sreenivasan-yeung-2005}
{\sc Schumacher, J., Sreenivasan, K.R. \& Yeung, P.K.} 2005 Very fine
  structures in scalar mixing. {\em J. Fluid Mech.\/} {\bf 531}, 113--122.

\bibitem[Schwertfirm \& Manhart(2007)]{schwertfirm-manhart-2007}
{\sc Schwertfirm, F. \& Manhart, M.} 2007 D{NS} of passive scalar transport in
  turbulent channel flow at high {S}chmidt numbers. {\em Int. J. Heat Fluid
  Flow\/} {\bf 28}, 1204--1214.

\bibitem[Shraiman \& Siggia(2000)]{shraiman-siggia-2000}
{\sc Shraiman, B.I. \& Siggia, E.D} 2000 Scalar turbulence. {\em Nature\/} {\bf
  405}, 639--646.

\bibitem[Sreenivasan(1991)]{sreenivasan-1991}
{\sc Sreenivasan, K.R.} 1991 On local isotropy of passive scalars in turbulent
  shear flows. {\em Proceedings of the Royal Society: Mathematical and Physical
  Sciences\/} {\bf 434}~(1890), 165--182.

\bibitem[Sreenivasan \& Antonia(1997)]{sreenivasan-antonia-1997}
{\sc Sreenivasan, K.R. \& Antonia, R.A.} 1997 The {P}henomenology of
  {S}mall-scale {T}urbulence. {\em Annu. Rev. Fluid Mech.\/} {\bf 29},
  435--472.

\bibitem[Wang {\em et~al.\/}(1999)Wang, Chen \&
  Brasseur]{wang-chen-brasseur-1999}
{\sc Wang, L.-P., Chen, S. \& Brasseur, J.G.} 1999 Examination of hypotheses in
  the {K}olmogorov refined turbulence theory through high-resolution
  simulations. {P}art 2. {P}assive scalar field. {\em J. Fluid Mech.\/} {\bf
  400}, 163--197.

\bibitem[Warhaft(2000)]{warhaft-2000}
{\sc Warhaft, Z.} 2000 Passive {S}calars in {T}urbulent {F}lows. {\em Annu.
  Rev. Fluid Mech.\/} {\bf 32}, 203--240.

\bibitem[Watanabe \& Gotoh(2007)]{watanabe-gotoh-2007b}
{\sc Watanabe, T. \& Gotoh, T.} 2007 Inertial-range intermittency and accuracy
  of direct numerical simulation for turbulence and passive scalar turbulence.
  {\em J. Fluid Mech.\/} {\bf 590}, 117--146.

\bibitem[Yeung {\em et~al.\/}(2005)Yeung, Donzis \&
  Sreenivasan]{yeung-donzis-sreenivasan-2005}
{\sc Yeung, P.K., Donzis, D.A. \& Sreenivasan, K.R.} 2005 High
  {R}eynolds-number simulation of turbulent mixing. {\em Phys. Fluids\/} {\bf
  17}, 081703.

\end{thebibliography}
\end{document}